\documentclass[traditabstract]{aa} 

\usepackage[fleqn]{amsmath}
\usepackage{amssymb}
\usepackage{textcomp}
\usepackage{graphicx}
\usepackage{txfonts}
\usepackage{natbib}
\usepackage{url}
\usepackage{multirow}
\usepackage{subfigure}
\usepackage[usenames,dvipsnames]{xcolor}

\newcommand{\ab}[1]{x({\rm #1})}
\newcommand{\op}[1]{{\rm OPR(#1)}}

\newcommand{\phh}{{\rm p\mathchar`- H_{2}}}
\newcommand{\ohh}{{\rm o\mathchar`- H_{2}}}

\newcommand{\react}[4]{{\rm #1} + {\rm #2} \rightarrow {\rm #3} + {\rm #4}}

\newcommand{\reacteq}[4]{{\rm #1} + {\rm #2} \rightleftharpoons {\rm #3} + {\rm #4}} 

\newcommand{\hdo}{{\rm HDO/H_{2}O}}
\newcommand{\ddo}{{\rm D_{2}O/HDO}}

\newcommand{\prob}[2]{P_{{\rm #1},\,#2}}
\newcommand{\fr}[2]{N_{{\rm #1},\,#2}}

\begin{document}

\title{Reconstructing the history of water ice formation from HDO/H$_2$O and D$_2$O/HDO ratios in protostellar cores}
\titlerunning{Reconstructing the history of water ice formation}
\authorrunning{Furuya et al.}

\author{K. Furuya\inst{\ref{inst1}} 
\and  E. F. van Dishoeck\inst{\ref{inst1},\ref{inst2}}
\and Y. Aikawa\inst{\ref{inst3}}}

\institute{Leiden Observatory, Leiden University, P.O. Box 9513, 2300 RA, The Netherlands\\
\email{furuya@strw.leidenuniv.nl}\label{inst1} 
\and Max-Planck-Institut f\"ur Extraterrestrische Physik, Giessenbachstrasse, Garching, Germany\label{inst2}
\and Center for Computer Sciences, University of Tsukuba, Tsukuba 305-8577, Japan\label{inst3}}

 
\abstract
{
Recent interferometer observations have found that the $\ddo$ abundance ratio is higher than 
that of $\hdo$ by about one order of magnitude in the vicinity of low-mass protostar NGC 1333-IRAS 2A, where water ice has sublimated.
Previous laboratory and theoretical studies show that the $\ddo$ ice ratio should be lower than the $\hdo$ ice ratio, 
if HDO and D$_2$O ices are formed simultaneously with H$_2$O ice.
In this work, we propose that the observed feature, $\ddo > \hdo$, is a natural consequence of chemical evolution 
in the early cold stages of low-mass star formation: 
1) majority of oxygen is locked up in water ice and other molecules in molecular clouds, where water deuteration is not efficient, and
2) water ice formation continues with much reduced efficiency in cold prestellar/protostellar cores, 
where deuteration processes are highly enhanced due to the drop of the ortho-para ratio of H$_2$, the weaker UV radiation field, etc.
Using a simple analytical model and gas-ice astrochemical simulations tracing the evolution from the formation of molecular clouds to protostellar cores, 
we show that the proposed scenario can quantitatively explain the observed $\hdo$ and $\ddo$ ratios.
We also find that the majority of HDO and D$_2$O ices are likely formed in cold prestellar/protostellar cores
rather than in molecular clouds, where the majority of H$_2$O ice is formed.
This work demonstrates the power of the combination of the $\hdo$ and $\ddo$ ratios as a tool to reveal the past history of 
water ice formation in the early cold stages of star formation and when the enrichment of deuterium in the bulk of water occurred.
Further observations are needed to explore if the relation, $\ddo > \hdo$, is common in low-mass protostellar sources.
}

\keywords{astrochemistry --- ISM: molecules --- ISM: clouds --- Stars: protostars}

\maketitle

\section{Introduction}
\label{sec:intro}
The degree of deuterium fractionation in molecules in general, and that of water in particular, depends on its formation environments.
This characteristic allows us to gain insights into the water trail from its formation in molecular clouds
to, ultimately, the delivery to planets by comparing the deuterium fractionation 
in objects at different evolutionary stages \citep[e.g., recent reviews by][]{ceccarelli14,vandishoeck14}.

It is well established that water is formed on grain surfaces in molecule clouds.
At dust temperatures lower than 100-150 K, water is predominantly present on the surfaces of dust grains as ice \citep{fraser01}.
There has been no clear detection of HDO ice in the interstellar medium (ISM), 
and the deuteration of water ice is not well-constrained \citep[$\hdo < (2-5) \times 10^{-3}$;][]{dartois03,parise03}.
Instead, there have been numerous observational studies on the deuteration of water vapor, 
which may reflect that of water ice through thermal and non-thermal desorption of ice.

In particular, recent interferometer and single-dish observations have quantified 
the degree of deuteration of water vapor in four low-mass protostellar sources, 
IRAS 16293-2422, NGC 1333-IRAS 2A, IRAS 4A, and IRAS 4B \citep{jorgensen10,liu11,coutens12,coutens13,coutens14a,persson13,persson14,taquet13}.
The interferometer observations provide the emission from the inner hot regions ($T > 100$ K),
where water ice is sublimated, while the single-dish observations including the $Herschel$ Space Observatory 
provide the integrated emission from larger spacial scales \citep{jorgensen10,coutens12}.
The combination of the two types of the observations have revealed that 
1) the gaseous $\hdo$ ratio in the inner hot regions ($\sim$10$^{-3}$) is lower than that in the cold outer envelopes by more than one order of magnitude, and
2) the gaseous $\ddo$ ratio is much higher than the $\hdo$ ratio ($\sim$10$^{-2}$ versus $\sim$10$^{-3}$) in the inner hot regions 
of NGC 1333-IRAS 2A \citep{coutens14a} and at least one other source (A. Coutens 2015, private communication).


Water deuteration in low-mass protostellar sources has also been studied using physical and astrochemical models 
\citep[e.g., recent work by][]{aikawa12b,taquet14,wakelam14}.
These studies adopt one-dimensional gravitational collapse models, which describe the physical evolution of 
collapsing prestellar cores to protostellar sources, 
with detailed gas-ice chemical networks.
The models successfully reproduce observed feature $\#1$, the radial gradient of the $\hdo$ ratio,  
by gas-phase ion-neutral chemistry in the outer cold regions and sublimation of ice in the inner hot regions.
The models, however, tend to overpredict the $\hdo$ ratio in the inner hot regions by a factor of several or more
compared to the observations.
Furthermore, in contrast to the observed feature $\#2$, 
all models predict that the gaseous $\ddo$ ratio in the inner hot regions is lower than or comparable to the $\hdo$ ratio.
\citet{coutens14a} have proposed that either there is something missing in the current understanding of deuterium chemistry on icy grain surfaces, or 
that water formation at high temperatures ($T_{\rm gas} > 200$--300 K) thorough reactions O + H$_2$ $\rightarrow$ OH + H 
and OH + H$_2$ $\rightarrow$ H$_2$O + H play a role in the inner quiescent regions, following sublimation of ice.
In the latter case, the $\ddo$ ratio reflects that in ice, while the $\hdo$ ratio is diluted by the additional formation of H$_2$O vapor, 
so that the $\ddo$ ratio can be higher than the $\hdo$ ratio.
However, it requires that a large amount of oxygen is in atomic form rather than in molecules in the high density inner regions.

In this paper, we propose an alternative scenario that can account for the higher $\ddo$ ratio compared with $\hdo$ 
by the combination of cold deuterium chemistry and sublimation of ice without the need for the enhanced H$_2$O formation in hot gas. 
This paper is organized as follows.
In Section \ref{sec:scenario}, the proposed scenario is presented through a simple analytical model.
In Section \ref{sec:model}, we simulate the gas-ice chemical evolution in star-forming cores with a numerical model to verify the scenario.
We briefly discuss molecular oxygen and the ortho-to-para ratio of H$_2$, the deuteration of methanol, 
and thermally induced H-D exchange reactions in ice in Section \ref{sec:discussion}.
Our findings are summarized in Section \ref{sec:conclusion}.

\section{Scenario}
\label{sec:scenario}
We assume that both the $\hdo$ and $\ddo$ gas ratios in the inner hot regions around protostars reflect those in ice.
In this subsection we denote the $\hdo$ ratio as $f_{\rm D1}$, while we denote the $\ddo$ ratio as $f_{\rm D2}$.

We first note that previous laboratory and theoretical studies show that $f_{\rm D2}$ should be lower than $f_{\rm D1}$, 
if H$_2$O, HDO, and D$_2$O ices are formed via grain surface reactions at the same time.
Let us assume that they all are formed via sequential reactions of atomic hydrogen/deuterium with atomic oxygen.
These reactions have no activation energy barrier \citep{allen77}.
If the surface reactions distribute deuterium statistically (or, in other words, mass-independently), 
the following relation holds \citep{rodgers02}: 
\begin{align}
[f_{\rm D2}/f_{\rm D1}]_{\rm statistic} = 0.25. \label{eq:f1f2_statistic}
\end{align}
Laboratory experiments have demonstrated that there are other formation pathways of water ice:
sequential surface reactions initiated by reactions of O$_2$/O$_3$ with atomic hydrogen \citep{ioppolo08,miyauchi08,mokrane09}, 
and the reaction OH + H$_2$ $\rightarrow$ H$_2$O + H \citep{oba12}.
These pathways include reactions with activation energy barriers, and thus proceed through quantum tunneling \citep{oba12,oba14}.
The barrier-mediated reactions favor hydrogenation over deuteration, because deuterium is twice heavier than hydrogen \citep[][]{oba14}.
In addition, once water is formed, it does not efficiently react with atomic deuterium to be deuterated at low temperatures 
unlike formaldehyde and methanol \citep[][]{nagaoka05}.
At warm temperatures ($>$70 K), H-D exchange reaction, ${\rm H_2O} + {\rm D_2O} \rightarrow {\rm 2HDO}$, is thermally activated, 
and the $f_{\rm D2}/f_{\rm D1}$ ratio can be lowered \citep[e.g.,][see also Section \ref{sec:hd_exchange}]{lamberts15}.
Taken together, 0.25 is the upper limit of the $f_{\rm D2}/f_{\rm D1}$ ratio.

The above constraint, $f_{\rm D2}/f_{\rm D1} \le 0.25$, can be directly applied to the compositions of ice in the ISM, 
if H$_2$O, HDO, and D$_2$O ices are predominantly formed simultaneously \citep{butner07}.
However, in the sequence of star formation, they are not necessarily formed at the same evolutionary stage \citep{dartois03}.
The onset of water ice mantle formation requires a threshold extinction, above which 
the photodesorption rate of water ice is lower than the formation rate of water ice and its precursors \citep[e.g.,][]{tielens05}.
The H$_2$O ice formation rate decreases with time as elemental oxygen is locked into O-bearing molecules.
Later in the evolution, at higher extinction and densities, 
CO freezes out and the products of its hydrogenation, such as formaldehyde (H$_2$CO) and methanol (CH$_3$OH), 
are thought to be the main constituent of the outer layers of the ice mantle \citep[e.g.,][]{pontoppidan06,oberg11}.
On the other hand, the formation rates of HDO and D$_2$O ices do not necessarily decrease together with that of H$_2$O ice;
deuterium fractionation is more efficient at later times, as CO is frozen out, 
the ortho-to-para nuclear spin ratio of H$_2$ ($\op{H_2}$) decreases, 
and interstellar UV radiation is heavily shielded \citep[e.g.,][and references therein]{caselli12}.

\begin{figure}
\resizebox{\hsize}{!}{\includegraphics{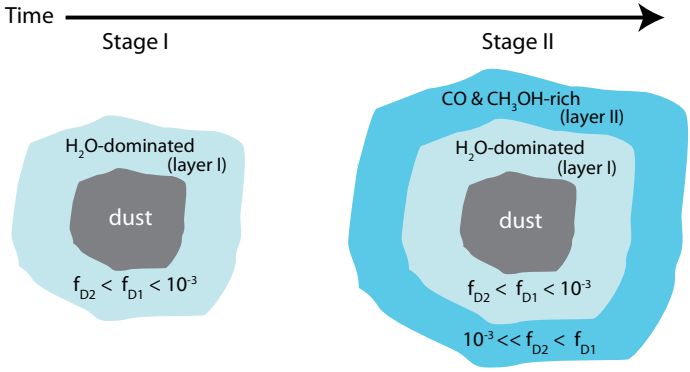}}
\caption{Schematic view of our scenario and the layered ice structure.
Stage I) The main formation stage of H$_2$O ice. Water deuteration is not efficient, $f_{\rm D2} < f_{\rm D1} < 10^{-3}$.
The majority of oxygen is locked in O-bearing molecules in this stage. 
Stage II) CO/CH$_3$OH-rich outer ice layers are formed, while the formation of water ice continues with much reduced efficiency compared to Stage I.
Nevertheless, the formation of HDO and D$_2$O ices is more efficient than in Stage I, due to the enhanced deuteration processes, $10^{-3} \ll f_{\rm D2} < f_{\rm D1}$. 
}
\label{fig:scenario}
\end{figure}

Infrared observations show that H$_2$O ice starts to become abundant in molecular clouds 
above a threshold line of sight visual extinction, depending on environments, e.g., $\sim$3 mag 
for Taurus dark clouds \citep[][and references therein]{whittet93,boogert15}.
The scenario we propose is that, in contrast to H$_2$O ice, HDO and D$_2$O ices are mainly formed 
at the later stages of star formation, i.e., in cold prestellar and protostellar cores, 
where deuterium fractionation processes would be more efficient than in ambient molecular clouds.
Then, considering the layered structure of ice mantles, HDO and D$_2$O are mainly present in the CO/CH$_3$OH-rich outer layers,
rather than the H$_2$O-dominated inner layers.
Assuming complete sublimation of the ice mantles in the inner hot regions of low-mass protostars, 
the gaseous $\ddo$ ratio directly reflects that in the outer layers of the ice mantles,
while the gaseous $\hdo$ ratio is much lower than that in the outer layers due to the dilution by the abundant H$_2$O in the inner layers.
Thus, the gaseous $\ddo$ ratio can be higher than the $\hdo$ ratio in the vicinity of protostars.
A schematic view of the layered ice structure in our scenario is shown in Figure \ref{fig:scenario}.
Our scenario is essentially similar to that proposed by \citet{dartois03} for HDO enhancement.
They suggested that water ice is formed without deuterium enrichment, followed by the additional water formation with 
high levels of deuterium fractionation at later times.
The motivation behind their scenario was to explain the upper limits of the icy $\hdo$ ratio in the cold outer envelope of protostars, 
which are lower than the D/H ratios of other gaseous species.

\begin{figure}
\resizebox{\hsize}{!}{\includegraphics{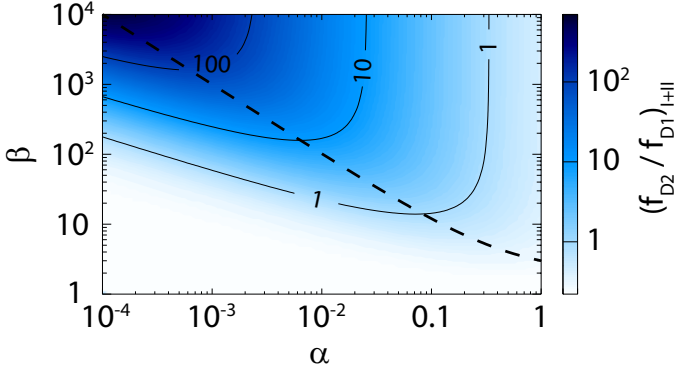}}
\caption{The $[f_{\rm D2}/f_{\rm D1}]_{\rm I+II}$ ratio as a function of $\alpha (=\fr{O}{\rm II}/\fr{O}{\rm I})$ and 
$\beta (=\prob{HDO}{\rm II}/\prob{HDO}{\rm I})$ in the two stage model given by Equation (\ref{eq:f2_f1_final}) with $q=0.25$. 
Above the dashed line ($\alpha\beta \gtrsim 1$), $[f_{\rm D1}]_{\rm I+II}$ 
is larger than $[f_{\rm D1}]_{\rm I}$ by a factor of two or more.
See the text for more information.}
\label{fig:f2_f1}
\end{figure}

To get a rough idea of how high the $f_{\rm D2}/f_{\rm D1}$ ratio can be in our scenario 
and to investigate the critical parameters, let us consider a two stage model (or a two layer ice model). 
We denote the total amount of oxygen locked into H$_2$O, HDO or D$_2$O ices at each stage (or in each layer) $k$, where $k=$ I or II, as $\fr{O}{k}$.
We denote the fraction of oxygen locked into X$_2$O ice, where X is H or D, as $\prob{X_2O}{k}$.
Then, $\prob{X_2O}{k}\fr{O}{k}$ represents the amount of X$_2$O ice formed in the stage $k$.
We can express $f_{\rm D1}$ and $f_{\rm D2}$ in the bulk of ice mantle as follows:
\begin{align}
[f_{\rm D1}]_{\rm I+II} = \sum_{k={\rm I,\,II}}\prob{HDO}{k}\fr{O}{k}\bigg/ \sum_{k={\rm I,\,II}}\prob{H_2O}{k}\fr{O}{k}, \\
[f_{\rm D2}]_{\rm I+II} = \sum_{k={\rm I,\,II}}\prob{D_2O}{k}\fr{O}{k} \bigg/ \sum_{k={\rm I,\,II}}\prob{HDO}{k}\fr{O}{k}.
\end{align}
It would be reasonable to assume $\prob{H_2O}{k}$ $\sim$ 1.
Motivated by Equation (\ref{eq:f1f2_statistic}), we introduce a free parameter $q$ 
as $\prob{D_2O}{k}/\prob{HDO}{k} = q\prob{HDO}{k}/\prob{H_2O}{k}$ in the range of $0 < q \le 0.25$.
When $q=0.25$, deuterium is statistically distributed, i.e., the most optimistic case to obtain a high $f_{\rm D2}/f_{\rm D1}$.
Although $q$ can vary between the two stages, we assume it is constant for simplicity.
With those relations, the $f_{\rm D2}/f_{\rm D1}$ ratio in the bulk of the ice mantle is expressed as
\begin{align}
\left[ \frac{f_{\rm D2}}{f_{\rm D1}} \right]_{\rm I+II} \approx \frac{q(1+\alpha)(1+\alpha \beta^2)}{(1+\alpha\beta)^2}, \label{eq:f2_f1_final}
\end{align}
where $\alpha$ is $\fr{O}{\rm II}/\fr{O}{\rm I}$ and $\beta$ is $\prob{HDO}{\rm II}/\prob{HDO}{\rm I}$.
Figure \ref{fig:f2_f1} shows $[f_{\rm D2}/f_{\rm D1}]_{\rm I+II}$ as a function of $\alpha$ and $\beta$ with $q=0.25$.
It shows that the proposed scenario can lead to $[f_{\rm D2}/f_{\rm D1}]_{\rm I+II} \gg 1$.
The conditions to reproduce the observed values in NGC 1333-IRAS 2A, $\hdo \sim 10^{-3}$ and $(\ddo)/(\hdo) \sim 10$, are that
1) most of water ice is formed with $\hdo < 10^{-3}$ and that
2) an additional small amount ($\alpha < 0.1$) of highly fractionated ($\beta \gtrsim 100$) water ice is formed at later times.

\citet[][hereafter Paper I]{furuya15} studied the chemical evolution from \mbox{\ion{H}{i}} dominated clouds to denser molecular clouds, 
following the scenario that \mbox{\ion{H}{i}} dominated gas is swept up and accumulated by global accretion flows \citep[e.g.,][]{hartmann01,inoue12}. 
Their primary goal was to investigate the evolution of the $\op{H_2}$ and the $\hdo$ ratio without an arbitrary assumption concerning the initial $\op{H_2}$.
It was found that the $\hdo$ ratio in the bulk ice can be much lower than 10$^{-3}$ at the end of the main formation stage of H$_2$O ice, 
i.e., condition $\#1$ can be satisfied.
In the following, we conduct the gas-ice chemical simulations of collapsing prestellar cores to the formation of protostars 
in order to explore if condition $\#2$ is fulfilled.

Based on the two stage model, we find that there are two regimes at the same $[f_{\rm D2}/f_{\rm D1}]_{\rm I+II}$.
Above the dashed line of Figure \ref{fig:f2_f1}, where $\alpha\beta \gtrsim 1$, 
the difference between $[f_{\rm D1}]_{\rm I+II}$ and $[f_{\rm D1}]_{\rm I}$ is more than a factor of two, 
i.e., the $\hdo$ ratio established at the main formation stage of H$_2$O ice is mostly hidden
by water ice additionally formed at later times.
Below the dashed line, on the other hand, $[f_{\rm D1}]_{\rm I+II}$ and $[f_{\rm D1}]_{\rm I}$ are similar.
The former corresponds exactly to our scenario, where HDO and D$_2$O ices are predominantly formed 
after the main formation stage of H$_2$O ice.
The latter is the case where H$_2$O and HDO ices are formed together, while D$_2$O ice is mainly formed after the main formation stage of H$_2$O and HDO ices.
For the same $[f_{\rm D2}/f_{\rm D1}]_{\rm I+II}$, the latter case requires more significant freeze out of oxygen, 
especially atomic oxygen and CO; 
water ice can be formed from CO gas through $\react{CO}{He^+}{C^+}{O}$.
Which regime is more likely in the ISM, or in other words, 
does the $\hdo$ ratio in the hot gas directly reflect the $\hdo$ ice ratio at the main formation stage of H$_2$O ice?
Numerical simulations are needed to answer the question.


\section{Numerical Simulation}
\label{sec:model}
\subsection{Model Description}

\begin{table*}
\caption{Initial Abundances of Selected Species with respect to Hydrogen Nuclei.}
\label{table:init}
\centering
\begin{tabular}{ccccccc}
\hline\hline
Species  &  MC1  &  MC2  &  MC3 &  AT1 & AT2 & AT4\\
\hline
$\ohh$            &2.4(-2) &1.1(-3) &2.4(-4) & 4.5(-2) & 5.0(-3) & 5.0(-5) \\
$\phh$            &4.8(-1) &5.0(-1) &5.0(-1) & 4.5(-1) & 5.0(-1) & 5.0(-1) \\
HD                &1.4(-5) &1.4(-5) &1.1(-5) & 1.5(-5) & 1.5(-5) & 1.5(-5) \\
\mbox{\ion{H}{i}} &5.2(-4) &7.3(-5) &5.1(-5) & -       & -       & -      \\
\mbox{\ion{D}{i}} &7.6(-7) &9.5(-7) &2.3(-6) & -       & -       & -      \\
H$_2$O            &6.9(-9) &5.2(-9) &9.2(-9) & -       & -       & -      \\
HDO               &1.2(-12)&3.7(-11)&1.3(-9) & -       & -       & -      \\
D$_2$O            & --     &2.9(-13)&8.1(-11)& -       & -       & -      \\
iH$_2$O           &8.4(-5) &1.1(-4) &1.2(-4) & -       & -       & -      \\
iHDO              &1.1(-8) &1.3(-8) &3.4(-8) & -       & -       & -      \\
iD$_2$O           &6.9(-13)&7.1(-13)&1.8(-11)& -       & -       & -      \\
\mbox{\ion{O}{i}} &2.2(-5) &1.7(-6) &2.3(-7) &1.8(-4)  & 1.8(-4) & 1.8(-4)\\
O$_2$             &1.9(-9) &4.0(-10)&3.2(-10)& -       & -       & -      \\
iO$_2$            & -      & -      & -      & -       & -       & -      \\
CO                &6.1(-5) &3.6(-5) &5.1(-6) & -       & -       & -      \\
iCO               &1.2(-5) &3.1(-5) &4.7(-5) & -       & -       & -      \\
\hline
\end{tabular}
\tablefoot{$a(-b)$ means $a\times10^{-b}$.
$\ohh$ indicates ortho-H$_2$, while $\phh$ indicates para-H$_2$. 
iX indicates species X in the bulk ice mantle. 
- indicates that abundances are less than 10$^{-13}$.
}
\end{table*}

We simulate water deuteration from a prestellar core to a protostellar core
adopting one-dimensional radiation hydrodynamics simulations of \citet{masunaga00}. 
Initially the core has an isothermal hydrostatic structure with a fixed outer boundary of $4\times10^4$ AU from the core center.
The total mass of the core is 3.9 $M_{\odot}$, which is greater than the critical mass for gravitational instability.  
The protostar is born at $2.5 \times 10^5$ yr after the beginning of the collapse, 
corresponding to 1.4$t_{\rm ff}$, where $t_{\rm ff}$ is the free-fall time of the initial central density 
of hydrogen nuclei $\sim$6$\times10^4$ cm$^{-3}$.
After the birth of the protostar, the model further follows the physical evolution for $9.3 \times 10^4$ yr.

Fluid parcels are traced in the hydrodynamics simulation, 
and we perform gas-ice chemical simulations along the stream lines to obtain the radial distribution of molecules 
in the protostellar envelope.
This approach is the same as \citet{aikawa12b}.
For simplicity, we set the temperature to be 10 K when the temperature in the original data is lower than 10 K.
We adopt a rate equation method and the chemistry is described by a three-phase model, 
which consists of gas, a chemically active icy surface, and inert ice mantles \citep{hasegawa93b}.
The top four monolayers of ice mantles are assumed to be chemically active, following \citet{vasyunin13}.
We refer to all of the layers including both the ice surface and the inert ice mantle as the bulk ice mantle.
We take into account gas-phase reactions, interaction between gas and (icy) grain surface, and surface reactions.
For non-thermal desorption processes, we consider stochastic heating by cosmic-rays \citep{hasegawa93a}, 
photodesorption \citep{westley95}, and chemical desorption \citep{garrod07}.
Our chemical reaction network is originally based on \citet{garrod06}.
The network has been extended to include high-temperature gas-phase reactions from \citet{harada10}, 
mono, doubly, and triply deuterated species \citep{aikawa12b,furuya13}, 
and nuclear spin states of H$_2$, H$_3^+$, and their isotopologues \citep[][]{hincelin14}.
The rate coefficients for the H$_2$ + H$_3^+$ system are taken from \citet{hugo09}.
More details can be found in Paper I.


We consider six sets of the initial abundances for the collapse model, which are summarized in Table \ref{table:init}.
In the sets labeled `MC', the molecular abundances are adopted from Paper I.
In Paper I, a one-dimensional shock model \citep{bergin04,hassel10} was used to study the chemical evolution during the formation
and growth of a molecular cloud via the accumulation of \mbox{\ion{H}{i}} gas.
Note that the evolution of the molecular cloud is dominated by ram pressure due to the accretion flow rather than self-gravity,
which is in contrast with our collapse model.
The time it takes for the column density of the cloud to reach $A_V=1$ mag is 
$\sim$4 Myr ($A_V$/1 mag)($n_0$/10 cm$^{-3}$)$^{-1}$ ($v_0$/15 km s$^{-1}$)$^{-1}$,
where $n_0$ and $v_0$ are the preshock \mbox{\ion{H}{i}} gas density and velocity of the accretion flow, respectively.
The density and temperature of the molecular cloud are $\sim$10$^4$ cm$^{-3}$ and 10-15 K, respectively.
We adopt the abundances when the column density of the molecular cloud 
reaches 1 mag (MC1), 2 mag (MC2), or 3 mag (MC3).
In all three sets, H$_2$O ice is abundant ($\ab{H_2Oice} \sim 10^{-4}$, where $x(i)$ is the abundance of species $i$ with respect 
to hydrogen nuclei) and the $\hdo$ ice ratio is $\sim$10$^{-4}$.
On the other hand, the abundances of atomic oxygen and CO in the gas phase, and the $\op{H_2}$ vary by orders of magnitude among the three sets; 
all three values decrease with increasing the column density of the cloud.
H$_2$ rather than CO is the key regulator of deuterium chemistry driven by $\reacteq{H_3^+}{HD}{H_2D^+}{H_2}$, 
as long as $\op{H_2}/\ab{CO} \gtrsim 40$ at temperatures of $\lesssim$20 K (Paper I).
This condition is fulfilled in all three sets.
The parameters $\alpha$ and $\beta$ discussed in Section \ref{sec:scenario} are related to 
($\ab{CO} + \ab{\mbox{\ion{O}{i}}}$) and the $\op{H_2}$, respectively.
In MC3, CO is largely frozen out, and its gas-phase abundance is only $5\times10^{-6}$.
The observations of CO isotopologues toward the low-mass prestellar cores have found that the CO abundance in the gas phase 
is lower than the canonical value, $\sim10^{-4}$, by a factor of around 10 \citep[e.g.,][]{crapsi05}. 
Then, the level of CO freeze out in MC3 is similar to that measured in prestellar cores.
Although the physical parameters vary continuously as a function of time in our numerical simulation, 
it can be considered as a two stage model, since we connect the two different physical models, i.e., cloud formation and core collapse.

In previous studies of deuterium chemistry in collapsing cores, species were initially assumed to be in atomic form 
except for H$_2$ and HD \citep[e.g.,][]{aikawa12b}.
For comparisons, we also perform chemical calculations in the collapsing core in which species are initially in atomic form 
except for H$_2$ and HD with the initial $\op{H_2}$ of 10$^{-1}$ (AT1), 10$^{-2}$ (AT2), or 10$^{-4}$ (AT4).

The visual extinction at the outer edge of the core is set to be 5 mag, 
being irradiated by the Draine field \citep{draine78}.
Paper I found that water ice deuteration can be significantly suppressed by interstellar UV radiation 
through the cycle of photodissociation and reformation of water ice, which efficiently removes deuterium from
water ice chemistry.
In this paper, our focus is placed on the water deuteration in the well-shielded regions.
The cosmic-ray ionization rate of H$_2$ is set to be $1.3\times10^{-17}$ s$^{-1}$, 
while the flux of FUV photons induced by cosmic-rays is set to be $3\times10^3$ cm$^{-2}$ s$^{-1}$.
In these settings, photochemistry is not important for water ice.

Before the onset of the collapse, we assume that the prestellar core keeps its hydrostatic structure for some time.
We consider three cases in which the elapsed time before the onset of the collapse is 
0 yr (labeled `A'), 10$^6$ yr (corresponding to $5.6t_{\rm ff}$, labeled `B'), or $3\times10^6$ yr ($16t_{\rm ff}$, `C').
Our fiducial model is MC2B; i.e., the initial molecular abundance is set by the cloud formation model at the epoch when the column density reaches 2 mag, 
and the duration of the static prestellar phase is 10$^6$ yr.

\subsection{Results}
\subsubsection{Fiducial model}
\begin{figure}
\resizebox{\hsize}{!}{\includegraphics{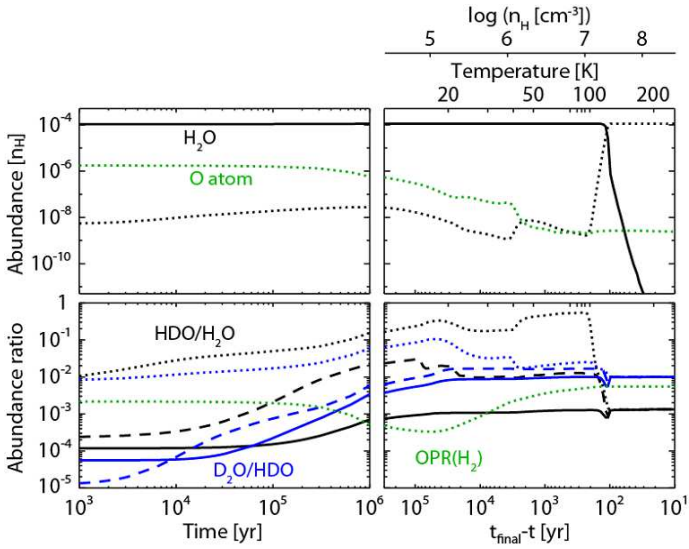}}
\caption{Temporal variations of molecular abundances (top) and abundance ratios (bottom) 
in the static phase (left) and during the collapse (right) in model MC2B in the fluid parcel, 
which reaches $R=5$ AU at the final time of the simulation.
Before the collapse begins, the temperature is 10 K and the number density of hydrogen nuclei ($n_{\rm H}$) is $2.3\times10^4$ cm$^{-3}$.
The horizontal axis of the right panels is set to be $t_{\rm final}-t$, 
where $t_{\rm final}$ represents the final time of the simulation and $t=0$ corresponds to the onset of the collapse.
The solid lines, the dashed lines, and the dotted lines represent molecules in the bulk ice mantle, molecules in the surface ice layers, 
and gaseous molecules, respectively.}
\label{fig:traj}
\end{figure}

\begin{figure}
\resizebox{\hsize}{!}{\includegraphics{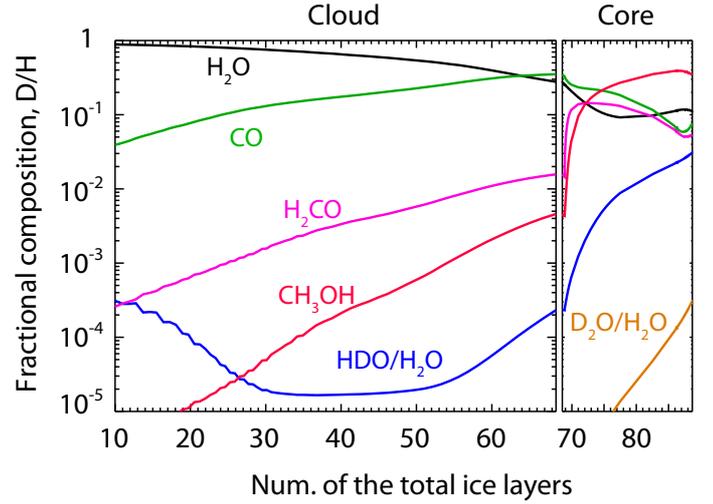}}
\caption{Fractional composition and D/H ratios in the surface ice layers as functions of the total number of icy layers in model MC2B.
The left panel represents the formation stage of a molecular cloud taken from Paper I, while the right panel represents the static and 
collapse stages at $t_{\rm final}-t < 8.3\times10^4$ yr, where dust temperature is 10 K, in the same fluid parcel shown in Figure \ref{fig:traj}.
The visual extinction for calculating photochemical rates increases from 2 mag at the end of the cloud formation stage to 5 mag at the prestellar core stage. 
The sudden increase leads to the sharp change of the fractional composition especially of H$_2$CO and CH$_3$OH in the surface ice layers.}
\label{fig:layer}
\end{figure}

Figure \ref{fig:traj} shows the temporal variation of molecular abundances and abundance ratios in the fiducial model (MC2B) 
along the stream line of a fluid parcel.
The fluid parcel is initially at 10$^4$ AU from the core center and finally reaches $5$ AU.
Figure \ref{fig:layer} shows the fractional composition and the D/H ratios in the active surface ice layers in the fluid parcel
as a function of the number of ice layers in total.
Note that surface layers become part of the inert ice mantle with the growth of ice.
Initially, most oxygen is locked up in CO and water ice with bulk ice ratios of $\hdo$ and $\ddo$ of $\sim$10$^{-4}$.
In the static phase and during the collapse at $T<20$ K, the environment is favorable for deuteration: 
the low temperature, 10-20 K, and the relatively low $\op{H_2}$, $\sim$10$^{-3}$, and the weak UV radiation field.
During those periods, corresponding to layers $\sim$70-90 in Figure \ref{fig:layer},
a small amount of additional water ice (the total abundance of $\sim5\times10^{-6}$) is formed with $\hdo > \ddo \gtrsim 10^{-2}$.
The source of oxygen to form the water ice is gaseous atomic oxygen and CO.
The additional water ice formation increases the concentrations of HDO and D$_2$O in the outer layers of the ice mantle, 
where CO and methanol are abundant as shown in Figure \ref{fig:layer}.
The $\ddo$ ratio in the bulk ice becomes similar to that in the surface layers with time, 
while the $\hdo$ ratio in the bulk ice remains much lower than that in the surface layers due to the abundant H$_2$O in the inert ice mantle.
The $\ddo$ ratio in the bulk ice becomes larger than the $\hdo$ ratio in $10^5$ yr in the static phase.

\begin{figure}
\resizebox{\hsize}{!}{\includegraphics{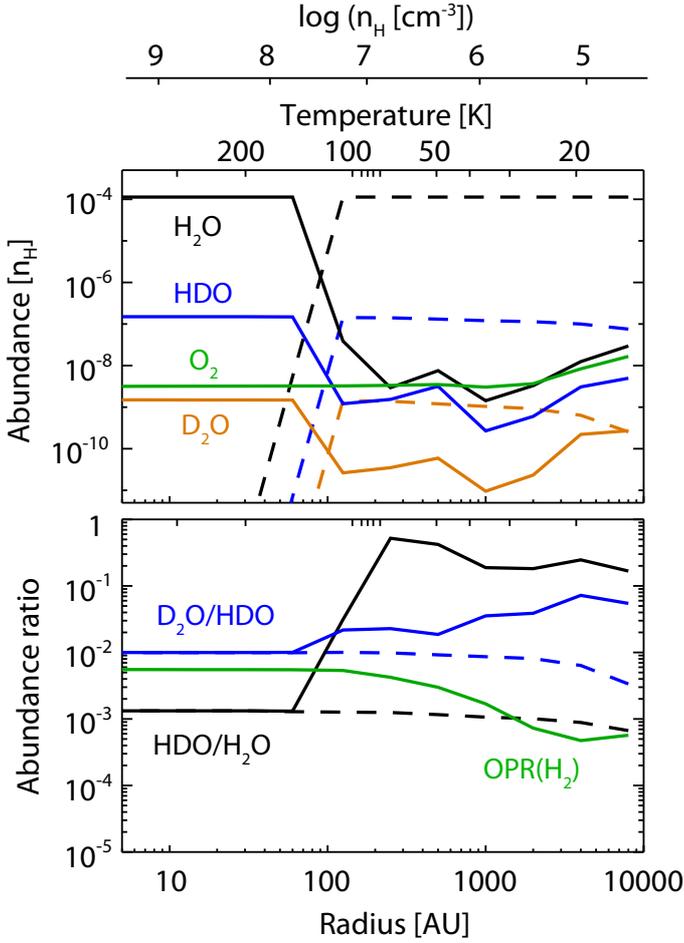}}
\caption{Radial profile of the molecular abundances (upper panel) and the abundance ratios of the molecules (lower panel) 
in model MC2B at $9.3\times10^4$ yr after the protostellar birth.
The labels at the top represents the temperature and density structures.
The solid lines represent gaseous molecules, while the dashed lines represent icy molecules in the bulk ice mantle.}
\label{fig:collapse}
\end{figure}

The top panel of Figure \ref{fig:collapse} shows radial profiles of the abundances of H$_2$O, HDO and D$_2$O 
at $9.3 \times 10^4$ yr after the protostellar birth, 
while the lower panel shows radial profiles of the $\hdo$ and $\ddo$ ratios.
The water sublimation radius, where the dust temperature is $\sim$150 K, is located at around 60 AU from the protostar;
water is mostly in the gas phase in the inner regions, while it predominantly exists as ice in the outer regions \citep{aikawa12b}.
At the water sublimation radius, both the gaseous $\hdo$ and $\ddo$ ratios drop via sublimation of ice.
At $T \gtrsim 150$ K, the gaseous ratios directly reflect those in the bulk ice.
At $T \lesssim 150$ K, the gaseous ratios are determined by ion-neutral chemistry in the gas phase; 
photodesorption is less important than the ion-neutral reactions, because of the weak UV radiation field dominated by the cosmic-ray induced UV.
We confirmed that the gaseous $\hdo$ and $\ddo$ ratios in the outer regions do not reflect either those in the bulk ice or those in the surface layers.
The model predicts $\hdo \sim 10^{-3}$ and $\ddo \sim 10^{-2}$ at $T \gtrsim 150$ K, 
which agree with the values measured in NGC 1333-IRAS 2A \citep{coutens14a}.
Our simulations demonstrate that the combination of cold deuterium chemistry and sublimation of ice can account for 
observed features $\#1$ and $\#2$, i.e., the radial gradient of the gaseous $\hdo$ ratio and 
the higher $\ddo$ than $\hdo$ ratios in the inner hot regions, simultaneously.

In the fiducial model, the $\hdo$ gas ratio at $T > 150$ K is higher than the initial $\hdo$ ice ratio by a factor of 11.
We denote this enhancement factor as $\varepsilon_{\rm D1}$.
The initial $\hdo$ ice ratio is mostly overshadowed by the small amount of highly fractionated water ice formed in the core.
To check the dependence of the result on the initial $\hdo$ ice ratio, we also run another two models,
in which the initial $\hdo$ ice ratio is artificially set to be 10$^{-5}$ (MC2B-L) and 10$^{-3}$ (MC2B-H), respectively.
Other details are the same as in the fiducial model.
In models MC2B-L and MC2B-H, $\varepsilon_{\rm D1}$ are 120 and 2, respectively.
The differences in the $\hdo$ ratios, on the other hand, at $T > 150$ K among the three models MC2B, MC2B-L, and MC2B-H 
are within a factor of two, in spite of the difference in the initial $\hdo$ ice ratio by two orders of magnitude.
Again it means that the initial $\hdo$ ice ratios are mostly overshadowed.

In our models, the main formation pathways of H$_2$O and HDO ices at $T < 20$ K are the surface reactions, 
OH + H$_2$ $\rightarrow$ H$_2$O + H and OD + H$_2$ $\rightarrow$ HDO + H, respectively.
Unexpectedly D$_2$O ice is formed via ion-neutral reactions in the gas phase followed by freeze out, instead of surface reactions.
To check whether our results depend on the main formation pathways of water ice,
we reran the fiducial model without the reaction OH + H$_2$ $\rightarrow$ H$_2$O + H and its deuterated analogues.
In this case, the main formation pathways of water ice are barrierless reactions of atomic hydrogen/deuterium with OH/OD.
We confirmed that the $\hdo$ and $\ddo$ ratios at $T>150$ K are similar to the fiducial case;
both ratios are enhanced by only $\sim$40 \%.

\subsubsection{Grid of models}
Table \ref{table:core} summarizes the results from our grid of the models with variations 
in the initial abundances and the elapsed time before the onset of the collapse.
In general, the models labeled MC predict a higher $\ddo$ ratio than $\hdo$ ratio at $T > 150$ K.
Therefore, specific conditions are not necessary to reproduce the observed ratios in IRAS 2A.
One exception is model MC1A, in which the initial $\op{H_2}$ is high ($\sim5\times10^{-2}$) and the collapse begins immediately.
The time required for the $\op{H_2}$ to reach steady state is longer than the free-fall timescale \citep{flower06}; 
the $\op{H_2}$ is higher than 10$^{-2}$ during the simulation, which reduces the efficiency of the overall deuteration processes, 
and prevents the enrichment of deuterated water in ice. 

The enhancement factor of the $\hdo$ ratio, $\varepsilon_{\rm D1}$, is also listed in Table \ref{table:core}.
In general, $\varepsilon_{\rm D1}$ is larger than 2 in models labeled MC.
Therefore, the $\hdo$ ice ratio established at the main formation phase of H$_2$O ice is mostly hidden 
by the additional formation of highly deuterated water ice at later times.

The models labeled AT tend to predict lower $\ddo$ ratios than $\hdo$ ratios at $T > 150$ K, 
in agreement with previous numerical studies \citep[e.g.,][]{aikawa12b}.
In these models, the gradient of deuterium fractionation seems not to be large enough.
This is due to the fact that the initial $\op{H_2}$, which is treated as a free parameter, controls the overall deuteration processes, 
again because the time required for the $\op{H_2}$ to reach steady state 
is longer than the free-fall timescale and the freeze-out timescale, i.e., the formation timescale of ice \citep{flower06,taquet14}.
Note that the steady state value of the $\op{H_2}$ for the initial dense core condition is of the order of 10$^{-4}$.
With the low initial $\op{H_2}$ of 10$^{-4}$, deuterium fractionation is already efficient at the main formation stage of H$_2$O ice,
while with the high initial $\op{H_2}$ of $\ge$10$^{-2}$, the efficiency of the overall deuterium fractionation is reduced during the 
cold prestellar core phase.
Among the models labeled AT, only models with a high initial $\op{H_2}$ and a very long static phase (AT1C and AT2C) predict 
higher $\ddo$ than $\hdo$ ratios.

\renewcommand\thefootnote{\alph{footnote}} 

\begin{table*}
\caption{Summary of Model Results}
\label{table:core}
\centering
\begin{tabular}{cccccccccc}
\hline\hline
\multirow{2}{*}{Model} & \multirow{2}{*}{H$_2$O\footnotemark[1]} & \raisebox{-0.75em}{$\dfrac{\rm HDO}{\rm H_2O}$}\footnotemark[2] & 
\raisebox{-0.75em}{$\dfrac{\rm D_2O}{\rm HDO}$}\footnotemark[2] & \raisebox{-0.75em}{$\dfrac{\rm D_2O/HDO}{\rm HDO/H_2O}$}\footnotemark[2] & 
\multirow{2}{*}{$\varepsilon_{\rm D1}$\footnotemark[3]} & \raisebox{-0.75em}{$\dfrac{\rm CH_3OH}{\rm H_2O}$}\footnotemark[2] &
\raisebox{-0.75em}{$\dfrac{\rm CH_3OD}{\rm CH_3OH}$}\footnotemark[2] &
\multirow{2}{*}{O$_2$\footnotemark[4]}& \multirow{2}{*}{$\op{H_2}$\footnotemark[5]}\\
\hline
MC1A   & 9.8(-5) & 2.8(-4) & 2.4(-4) & 0.8 & 2.2 & 7.8(-2) & 3.4(-4) & 4.1(-7) & 3.1(-2)\\
MC1B   & 1.0(-4) & 6.0(-4) & 2.1(-3) & 3.5 & 4.6 & 2.2(-1) & 6.4(-4) & 7.6(-8) & 3.7(-3)\\
MC2A   & 1.1(-4) & 5.8(-4) & 4.1(-3) & 7.2 & 4.8 & 4.8(-2) & 4.0(-3) & 2.7(-8) & 1.5(-3)\\
MC2B   & 1.1(-4) & 1.3(-3) & 1.0(-2) & 7.5 & 11  & 1.5(-1) & 6.1(-3) & 1.6(-8) & 4.7(-4)\\
MC3A   & 1.2(-4) & 7.1(-4) & 1.3(-2) & 19  & 2.5 & 2.5(-2) & 1.1(-2) & 4.9(-9) & 3.4(-4)\\
MC3B   & 1.2(-4) & 9.5(-4) & 1.6(-2) & 17  & 3.4 & 4.3(-2) & 1.4(-2) & 1.9(-9) & 2.5(-4)\\
\hline
AT1B   & 1.2(-4) & 6.7(-4) & 2.5(-4) & 0.4 & -   & 7.5(-2) & 2.9(-4) & 7.8(-7) & 1.4(-2)\\
AT1C   & 1.2(-4) & 1.3(-3) & 1.0(-2) & 7.8 & -   & 1.2(-1) & 2.5(-3) & 9.3(-9) & 3.0(-4)\\
AT2B   & 1.2(-4) & 2.4(-3) & 1.3(-3) & 0.6 & -   & 7.4(-2) & 2.0(-3) & 7.8(-7) & 2.2(-3)\\
AT2C   & 1.2(-4) & 3.2(-3) & 6.3(-3) & 2.0 & -   & 1.2(-1) & 5.0(-3) & 9.3(-9) & 2.8(-4)\\
AT4B   & 1.1(-4) & 6.4(-3) & 2.5(-3) & 0.4 & -   & 7.2(-2) & 5.9(-3) & 7.8(-7) & 5.4(-4)\\
AT4C   & 1.2(-4) & 7.0(-3) & 4.2(-3) & 0.6 & -   & 1.2(-1) & 7.9(-3) & 9.3(-9) & 2.7(-4)\\
\hline
MC2B-L & 1.1(-4) & 1.2(-3) & 1.0(-2) & 8.8 & 120 & 1.5(-1) & 6.1(-3) & 1.6(-8) & 4.7(-4)\\
MC2B-H & 1.1(-4) & 2.1(-3) & 6.1(-3) & 2.8 & 2.1 & 1.5(-1) & 6.1(-3) & 1.6(-8) & 4.7(-4)\\

\hline
\end{tabular}
\tablefoot{Values at $9.3\times10^4$ yr after the protostellar birth.\\
\tablefoottext{a}{The abundance of H$_2$O gas with respect to hydrogen nuclei at $T>150$ K.}
\tablefoottext{b}{Abundance ratios at $T>150$ K.}
\tablefoottext{c}{The ratio of the gaseous $\hdo$ ratio at $T>150$ K to the initial $\hdo$ ice ratio.}
\tablefoottext{d}{The maximum abundance of O$_2$ gas with respect to hydrogen nuclei at $T<150$ K.}
\tablefoottext{e}{The minimum ortho-to-para nuclear spin ratio of H$_2$ at $T<150$ K.}
}
\end{table*}

\section{Discussion}
\label{sec:discussion}
\subsection{O$_2$ and ortho-para ratio of H$_2$}
In the simulations of the collapsing core, the critical parameters controlling the $\hdo$ and $\ddo$ ratios in the hot gas around
protostars are the $\op{H_2}$ and the amount of oxygen, i.e., atomic oxygen and CO, that is available for water ice formation.
Here we briefly compare our model predictions with the observationally derived O$_2$ abundance and $\op{H_2}$ toward low-mass protostars.
Although the abundance of atomic oxygen in the cold outer envelopes is hard to constrain via observations, 
we can partly verify the oxygen chemistry in our models by comparing the predicted O$_2$ abundance with that derived from observations.

\citet{yildiz13} derived an upper limit of O$_2$ gas abundance $\lesssim$10$^{-8}$ toward NGC 1333-IRAS 4A, 
assuming a constant abundance in the envelope.
\citet{brunken14} derived the $\op{H_2D^+}$ of $\sim$0.1 in the cold outer envelope of IRAS 16293-2422, 
which corresponds to an $\op{H_2}$ of $2\times10^{-4}$ in their best fit model.
In general, the models which predict a higher $\ddo$ than $\hdo$ ratio in the gas phase at $T > 150$ K reproduce
the observations of O$_2$ and $\op{H_2}$ better than the other models.
Our fiducial model, for example, is consistent with the upper limit of the O$_2$ abundance and the $\op{H_2}$ derived from the observations, 
as shown in Figure \ref{fig:collapse} and Table \ref{table:core}.
Unless a very long static phase ($16t_{\rm ff}$) is assumed,
the models labeled AT tend to overpredict the O$_2$ abundance, implying that the majority of oxygen should be locked up in molecules
such as water and CO before the prestellar core stage \citep{yildiz13,bergin00}.

In summary, our models labeled MC reasonably reproduce the observations of the water deuterium fractionation, 
the O$_2$ abundance, and the $\op{H_2}$, simultaneously.
This strengthens our scenario.
Note, however, that these observed values are measured in different sources, and 
it is unclear whether the higher $\ddo$ ratio compared with $\hdo$, the low O$_2$ abundance, and the low $\op{H_2}$ are 
general chemical features of low-mass protostellar sources.


\subsection{Water deuteration versus Methanol deuteration}
In the inner hot regions ($T \gtrsim 100$ K) of low-mass protostellar sources, formaldehyde and methanol show higher levels of deuterium fractionation than water.
For example, the CH$_3$OD/CH$_3$OH ratio \citep[$\gtrsim$10$^{-2}$;][]{parise06} is much higher than the $\hdo$ ratio \citep[$\sim$10$^{-3}$; e.g.,][]{persson14}.
This difference is thought to reflect the different epoch of their formation, i.e., water ice is formed 
in an earlier stage of star formation than formaldehyde and methanol ices \citep{cazaux11,taquet12}.
Note that the scenario implicitly assumes that H$_2$O and HDO ices are formed at the same evolutionary stage.

In our scenario, H$_2$O ice is mainly formed in molecular clouds, while HDO and D$_2$O ices are mainly formed at later times 
when CO and methanol are the main constituent of the surface layers of the ice.
In other words, H$_2$O ice is formed in an earlier stage of star formation than deuterated water, formaldehyde and methanol ices.
This naturally leads to the following relation for the abundance ratios, CH$_3$OD/CH$_3$OH $\sim$ D$_2$O/HDO $>$ H$_2$O/HDO, 
simply because methanol and deuterated water are formed at the similar epoch.
The CH$_3$OD/CH$_3$OH ratios at $T>150$ K in our grid of models are presented in Table \ref{table:core} with the CH$_3$OH/H$_2$O ratios.
Our fiducial model, for example, predicts CH$_3$OD/CH$_3$OH = $6\times10^{-3}$, D$_2$O/HDO = $1\times10^{-2}$, and HDO/H$_2$O = $1\times10^{-3}$,
which agree reasonably well with the observations.
Here we considered only the CH$_3$OD/CH$_3$OH ratio, since the ratio is determined by hydrogenation/deuteration of CO on a icy grain surface.
Other species, such as HDCO and CH$_2$DOH, are subject to the abstraction and substitution reactions \citep[][]{watanabe08}, 
and the situation is more complex \citep[see][for a detailed discussion]{taquet12}.

\subsection{Thermally induced H-D exchange reactions \label{sec:hd_exchange}}
Laboratory experiments have shown that thermally activated H-D exchanges between hydrogen-bonded molecules 
in mixed ices occur efficiently at warm temperatures of $\gtrsim$70 K \citep{ratajczak09,faure15,lamberts15}.
This type of reactions is not included in our chemical network.
In particular for the present study, \citet{lamberts15} experimentally studied the thermally activated H-D exchange reactions, 
\begin{align}
{\rm H_2O} + {\rm D_2O} \rightleftharpoons {\rm 2HDO}, \label{react:exchange} 
\end{align}
in mixed amorphous ices.
In equilibrium, the ratio of the $\ddo$ ratio to the $\hdo$ ratio is given as 
$k_{\ref{react:exchange}b}/k_{\ref{react:exchange}f}$, where $k_{\ref{react:exchange}b}$ and $k_{\ref{react:exchange}f}$
are the reaction rate coefficient of Reaction (\ref{react:exchange}) in the backward direction and the forward direction,
respectively.
Lamberts et al. found that the activation energy barrier of the forward reaction is $3840\pm125$ K, 
which is much smaller than the binding energy of water on a water substrate, 5700 K \citep{fraser01}.
Although the activation barrier for the backward reaction has not been measured in laboratory, 
the backward reaction is energetically less favorable than the forward reaction, 
i.e., likely $k_{\ref{react:exchange}b}/k_{\ref{react:exchange}f} < 1$ \citep{collier84,lamberts15}.
Taken together, the $\ddo$ ice ratio can be lowered during the warm-up of ices in in-falling protostellar envelopes \citep{lamberts15}.

To check the impact of the H-D exchange by Reaction (\ref{react:exchange}), 
we compared the timescale of forward reaction calculated by \citet{lamberts15} 
and the duration time of the warm temperatures in our collapse model, and found that the former is much shorter than the latter at $\gtrsim$80 K.
Then D$_2$O in ice would be largely lost prior to the sublimation of water ice,
if Reaction (\ref{react:exchange}) is as efficient in the ISM as in laboratory.
This challenges the D$_2$O observations by \citet{coutens14a}, who showed that the D$_2$O emission peak is located at 
the peak of the continuum emission as well as the H$_2^{18}$O and the HDO emissions.

A possible explanation for the discrepancy is the layered ice structure in the ISM.
\citet{ratajczak09} experimentally demonstrated that thermally activated H-D exchange between CD$_3$OD and H$_2$O
occurs efficiently in mixed ice at $T>120$ K, while the exchange does not occur in the case of 
segregated ice even at higher temperatures on a laboratory timescale.
As noted in \citet{ratajczak09}, this implies that the H/D exchange occurs only between closely interacting molecules in ice.
In the experiments by \citet{lamberts15}, mixed ices with the mixing ratio of (H$_2$O:D$_2$O)$\sim$(1:1) were used,
which ensures that each D$_2$O has at least one neighbour of H$_2$O.
In our scenario, however, D$_2$O is mainly present in the CO/CH$_3$OH-rich outer-layers, 
i.e., significant fraction of D$_2$O would not be neighbored by H$_2$O molecules, 
and it may be the limiting factor of the H/D exchange by Reaction (\ref{react:exchange}) in the ISM.


\section{Conclusion}
\label{sec:conclusion}
Recent interferometer observations have found that the $\ddo$ ratio is higher than the $\hdo$ ratio in the warm gas surrounding 
the low-mass protostar NGC 1333-IRAS 2A, where water ice has sublimated \citep{coutens14a}.
In this study, we have proposed that the observed feature is a natural consequence of chemical evolution in the early cold stages of low-mass star formation:
1) majority of oxygen is locked up in water ice and other molecules in molecular clouds, where water deuteration is not efficient, and
2) water ice formation continues with much reduced efficiency in cold prestellar/protostellar cores, where deuteration processes are highly enhanced.
Using a simple analytical model and gas-ice astrochemical simulations on the formation of molecular clouds (Paper I) and the gravitational collapse of dense cores, 
we have shown that the scenario can quantitatively explain the $\hdo$ ratio and the $\ddo$ ratio measured in IRAS 2A.
Our model predictions are consistent with the low abundance of O$_2$, the low $\op{H_2}$, and the CH$_3$OD/CH$_3$OH ratio measured in 
low-mass protostellar sources, which further supports the scenario.
We also found that the majority of HDO and D$_2$O ices are likely formed in cold prestellar/protostellar cores
rather than in molecular clouds, where the majority of H$_2$O ice is formed.
Considering the layered ice mantles, this implies that HDO and D$_2$O are predominantly present in the CO/CH$_3$OH-rich outer layers of ice mantles 
rather than in the H$_2$O-dominated inner layers.
The layered ice structure indicates that the gaseous $\hdo$ ratio produced by photodesorption 
$does$ $not$ reflect that of the bulk ice \citep[see also][]{taquet14}.

The present study demonstrates the power of the combination of the $\hdo$ and $\ddo$ ratios as a tool to reveal the past history of 
water ice formation in the early cold stages of star formation and when the enrichment of deuterium in the bulk of water occurred.
Further observations are desirable to investigate if the relation, $\ddo > \hdo$, is common in low-mass protostellar sources.

\begin{acknowledgements}
We thank Magnus V. Persson, Thanja Lamberts, Vianney Taquet, Catherine Walsh, and Maria N. Drozdovskaya for interesting discussions.
We also thank the referee for his/her comments.
Astrochemistry in Leiden is supported by the Netherlands Research School for Astronomy (NOVA), by a Royal Netherlands Academy of Arts
and Sciences (KNAW) professor prize, and by the European Union A-ERC grant 291141 CHEMPLAN.
K.F. is supported by the Research Fellowship from the Japan Society for the Promotion of Science (JSPS).
Numerical computations were in part carried out on PC cluster at Center for Computational Astrophysics, National Astronomical Observatory of Japan.
\end{acknowledgements}

\end{document}